\documentclass[10pt,twocolumn]{article}

\usepackage{authblk}
\usepackage{amsmath}
\usepackage{amssymb}
\usepackage{amsthm}

\newtheorem{definition}{Definition}
\usepackage{booktabs} 
\usepackage{subcaption}
\usepackage{graphicx}
\usepackage{algorithm}
\usepackage{algorithmic}
\usepackage{placeins}
\usepackage{mathtools}
\usepackage{url}
\usepackage{enumitem}
\usepackage{multirow}
\usepackage{rotating}
\usepackage{nicefrac}
\usepackage{epstopdf}

\begin{document}
\title{\textbf{DIMSpan - Transactional Frequent Subgraph Mining with Distributed In-Memory Dataflow Systems}}

\author{Andr\'{e} Petermann, Martin Junghanns and Erhard Rahm}

\affil{University of Leipzig \& ScaDS Dresden/Leipzig}
\affil{\texttt{[petermann,junghanns,rahm]@informatik.uni-leipzig.de}}

\date{}
\pagestyle{empty}

\maketitle

\begin{abstract}
Transactional frequent subgraph mining identifies frequent subgraphs in a collection of graphs. 
This research problem has wide applicability and increasingly requires higher scalability over 
single machine solutions to address the needs of Big Data use cases. 
We introduce DIMSpan, an advanced approach to frequent subgraph mining that utilizes the features provided by distributed in-memory dataflow systems such as  Apache Spark or Apache Flink. It determines the complete set of frequent subgraphs from arbitrary string-labeled directed multigraphs as they occur in social, business and knowledge networks. DIMSpan is optimized to runtime and minimal network traffic but memory-aware. An extensive performance evaluation on large graph collections shows the scalability of DIMSpan and the effectiveness of its pruning and optimization techniques. 
\end{abstract}

\begin{table*}

\caption{Glossary of symbols}
\label{tab:}

\begin{center}

\begin{tabular}{|ll|}
\hline
$G / v / e / P / m$ & graph / vertex / edge / pattern / embedding \\
$\mathcal{G} / V / E / \mathcal{P} / M$ & sets of $G / v / e / P / m$ \\
$\mathcal{F} / \mathcal{\mu}$ & set of frequent patterns / pattern-embeddings map \\
$\phi /\phi_w$ &pattern frequency / frequency within a partition \\
$?^k / \mathcal{G}_{i \in \mathbb{N}}$ & k-edge variant of $?$ / partition of a graph set \\
$C_{min}(P)$ & minimum DFS code of a pattern \\
$C^1(e)$ & minimum DFS code of an edge \\
$C^1(P)$ & first extension of a pattern's min. DFS code \\
\hline
\end{tabular}

\end{center}
\end{table*}

\section{Introduction}

Mining frequent structural patterns from a collection of graphs, usually referred to as \textit{frequent subgraph mining} (FSM), has found much research interest in the last two decades, for example, to identify significant patterns from chemical or biological structures and protein interaction networks \cite{jiang2013survey}. Besides these typical application domains, graph collections are generally a natural representation of partitioned network data such as knowledge graphs \cite{cyganiak2008n}, business process executions \cite{petermann2014biiig} or communities in a social network \cite{junghanns2016epgm}. We identified two requirements for FSM on such data that are not satisfied by existing approaches: First, such data typically describes directed multigraphs, i.e., the direction of an edge has a semantic meaning and there may exist multiple edges between the same pair of vertices. 
Second, single machine solutions will not be sufficient for big data scenarios where either input data volume as well as size of intermediate results can exceed main memory or achievable runtimes are not satisfying. 

An established approach to speed up or even enable complex computations on very large data volumes is data-centric processing on clusters without shared memory. 
The rise of this approach was strongly connected with the MapReduce \cite{dean2008mapreduce} programming paradigm, which has also been applied to the FSM problem \cite{hill2012iterative, lu2013efficiently, aridhi2014novel, lin2014large,  bhuiyan2015iterative}. However, none of the approaches provides support for directed multigraphs. Further on, MapReduce is not well suited for complex iterative problems like FSM as it leads to a massive overhead of disk access. 

In recent years, a new generation of advanced cluster computing systems like Apache Spark \cite{zaharia2012resilient} and Apache Flink \cite{carbone2015apache}, in the following denoted by \textit{distributed in-memory dataflow systems}, appeared. In contrast to MapReduce, these systems provide a larger set of operators and support holding data in main memory between operators as well as during iterative calculations. 

In this work, we propose DIMSpan, an advanced approach to distributed FSM based on this kind of system. Our contributions can be summarized as follows:
\vspace{-1mm}

\begin{itemize}[leftmargin=4mm]
\itemsep0.2em 
\item  We propose DIMSpan, the first approach to parallel FSM based on distributed in-memory dataflow systems (Section \ref{sec:algorithm}). It adapts all pruning features of the popular gSpan \cite{yan2002gspan} algorithm to the dataflow programming model. Further on, it supports directed multigraphs and its data structures are optimized to pruning and compression .
  \item We provide a comparison to existing MapReduce based approaches (Section \ref{sec:mr}) and show that DIMSpan not only requires fewer disk access but also shuffles less data over the network and can reduce the total number of expensive isomorphism resolutions to a minimum.
  
  \item We present results of experimental evaluations (Section \ref{sec:eval}) based on real and synthetic datasets to show the scalability of our approach as well as the runtime impact of single pruning and optimization techniques .
  \item Our implementation is practicable and works for arbitrary string-labeled graphs. We provide its source code to the community as part of the \textsc{Gradoop} framework \cite{petermann2016graph} under an Open Source licence. 
\end{itemize}
\vspace{-1mm}
In addition, we provide background knowledge and discuss related work in Section \ref{sec:background}. Finally,  we conclude and give a preview on future work in Section \ref{sec:conclusion}.

\vspace{-3mm}

\section{Background \& Related Work}
\label{sec:background}

In this section, we introduce the distributed dataflow programming model, define the frequent subgraph mining problem and discuss related work.

\begin{table*}
\centering

\caption{Selected Unary Tranformations}
\label{tab:unary}

\begin{tabular}{|lll|}
\hline
\textbf{Transf.} & \textbf{Signature}\hspace{5mm} & \textbf{Constraints} \\
\hline
\multicolumn{3}{|l|}{\textit{single element transformations}} \\
Filter    & $I, O \subseteq A$   & $O \subseteq I$\\
Map       & $I\subseteq A, O \subseteq B $   & $\left| I \right| = \left| O \right|$\\
Flatmap   & $I\subseteq A, O \subseteq B $   & -\\
MRMap     & $I \subseteq A \times B; O \subseteq C \times D$ & -   \\
\hline
\multicolumn{3}{|l|}{\textit{element group transformations}} \\
Reduce    & $I, O \subseteq A \times B$&$\left| I \right| \geq \left| O \right| \wedge \left| O \right| \leq \left| A \right|$\\
Combine    & $I, O \subseteq A \times B$&$\left| I \right| \geq \left| O \right| \wedge \left| O \right| \leq \left| A \times W \right|$\\
\hline
\end{tabular}

\vspace{2mm}
(I/O : input/output datasets, A..D : domains, W : worker threads)\\
\vspace{8mm}
\end{table*}


\begin{figure*}[t]
	\centering

	\caption{Example illustrations for a graph collection, a subgraph, a pattern lattice and embeddings.}
  \includegraphics[width=0.68\textwidth]{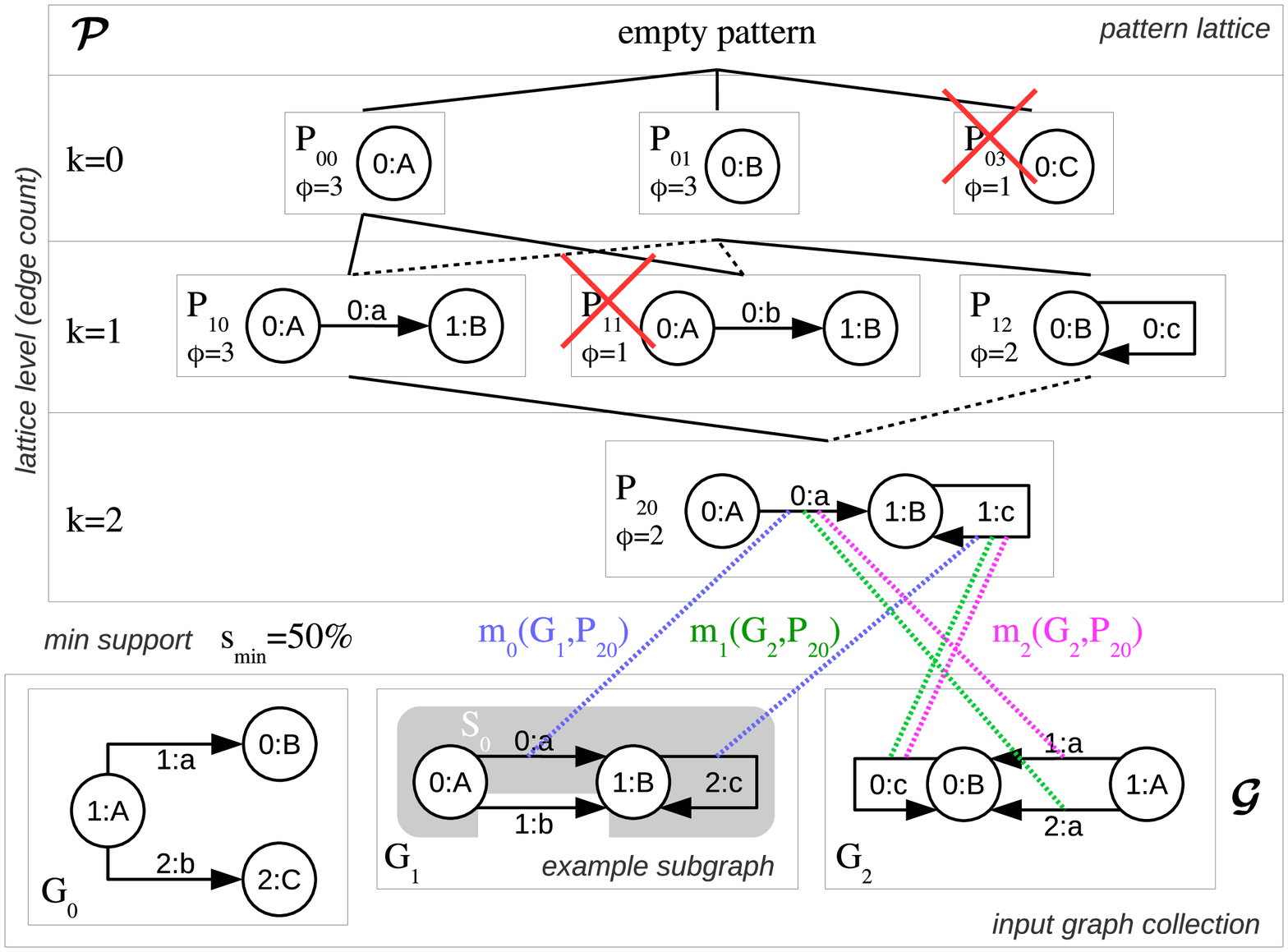}
\label{fig:lattice}

\end{figure*}

\subsection{Distributed Dataflow Model}

Distributed dataflow systems like MapReduce \cite{dean2008mapreduce}, Apache Spark \cite{zaharia2012resilient} or Apache Flink \cite{carbone2015apache} are designed to implement data-centric algorithms on shared nothing clusters without handling the technical aspects of parallelization. The fundamental programming abstractions are datasets and transformations among them. A \textit{dataset} is a set of data objects partitioned over a cluster of computers. A \textit{transformation} is an operation that is executed on the elements of one or two input datasets. The output of a transformation is a new dataset. Transformations can be executed concurrently on $W = \{w_0,w_1,..,w_n\}$ available \textit{worker threads}, where every thread executes the transformation on an associated dataset partition. 
There is no shared memory among threads.

Depending on the number of input datasets we distinguish \textit{unary} and \textit{binary} transformations. Table \ref{tab:unary} shows example unary tranformations. We further divide them into those transformations processing \textit{single elements} and those processing \textit{groups of elements}. All of the shown functions require the user to provide a \textit{transformation function} $\tau$ that needs to be executed for each element or group. A simple transformation is \textit{filter}, were $\tau$ is a predicate function and only those elements for which $\tau$ evaluates to true will be added to the output. Another simple transformation is \textit{map}, where $\tau$ describes how exactly one output element is derived from an input element. \textit{Flatmap} is similar to map but allows an arbitrary number of output elements. MapReduce provides only one single-element transformation (denoted by \textit{MRMap} in Table \ref{tab:unary}) which is a variant of flatmap that requires input and output elements to be key-value pairs. 

The most important element group transformation is \textit{reduce}. Here, input as well as output are key-value pairs and for each execution all elements sharing the same key are aggregated and $\tau$ describes the generation of a single output pair with the same key. Since input pairs with the same key may be located in different partitions they need to be \textit{shuffled} among threads which is typically causing network traffic among physical machines. If $\tau$ is associative (e.g. summation), an additional combine transformation can be used to reduce this traffic. \textit{Combine} is equivalent to reduce but skips shuffling, i.e., in the worst case one output pair is generated for each key and thread. Afterwards, these partial aggregation results can be passed to a reduce transformation.

As map and filter can also be expressed using MRMap, MapReduce and the new generation of \textit{distributed in-memory dataflow systems} (DIMS) like Spark and Flink have the same expressive power in terms of unary transformations. However, in the case of successive or iterative MRMap-reduce phases intermediate results need to be read from disk at the beginning and written to disk at the end of each phase. Thus, MapReduce is not well suited to solve iterative problems and problem-specific distributed computing models arose, for example, to process very large graphs \cite{mccune2015thinking}. In contrast, MapReduce and DIMS are general purpose platforms and not dedicated to a specific problem. However, DIMS support more complex programs including iterations, binary transformations (e.g., set operators like \textit{union} and \textit{join}) and are able to hold datasets in main memory during the whole program execution.

%

\subsection{Frequent Subgraph Mining}
\label{sec:problem}

Frequent subgraph mining (FSM) is a variant of frequent pattern mining \cite{aggarwal2014frequent} where patterns are graphs. There are two variants of the FSM problem. \textit{Single graph FSM} identifies patterns occurring at least a given number of times within a single graph, while \textit{graph transaction FSM} searches for patterns occurring in a minimum number of graphs in a collection. Our proposed approach belongs to the second setting. Since there exist many variations of this problem we first define our problem precisely before discussing related work and introducing our algorithm.


\pagebreak

\begin{definition}
\label{def:graph}\textsc{(Graph)}. 
Given two global label sets $\mathcal{L}_v, \mathcal{L}_e$, then a \textit{directed labeled multigraph}, in the following simply referred to as \textit{graph}, is defined to be a hextuple $G = \langle V,E, s, t, \lambda_v, \lambda_e \rangle$, where $V = \{v\}$ is the set of vertices (vertex identifiers), $E = \{e\}$ is the set of edges (edge identifiers), the functions $s : E \rightarrow V\ /\ t : E \rightarrow V$ map a \textit{source} and a \textit{target} vertex to a every edge and $\lambda_v : V \rightarrow \mathcal{L}_v\ /\ \lambda_e : E \rightarrow \mathcal{L}_e$ associate labels to vertices and edges. An edge $e \in E$ is \textit{directed} from $s(e)$ to $t(e)$. A multigraph supports loops and parallel edges.
\end{definition}

\begin{definition}
\label{def:subgraph}\textsc{(Subgraph).} 
Let $S, G$ be graphs then $S$ will be considered to be a \textit{subgraph} of $G$, in the following denoted by $S \sqsubseteq G$, if $S$ has subsets of vertices $S.V \subseteq G.V$ and edges $S.E \subseteq G.E$ and $\forall e \in S.E : s(e), t(e) \in S.V$ is true.
\end{definition}

\hspace{-4mm}
On the bottom of Figure \ref{fig:lattice}, a collection of directed multigraphs $\mathcal{G} = \{G_1, G_2,G_3\}$ and an example subgraph $S_0 \sqsubseteq G_1$ are illustrated. Identifiers and labels of vertices and edged are encoded in the format \texttt{id:label}, e.g., \texttt{1:A}.

\begin{definition}
\label{def:isomorphism}\textsc{(Isomorphism).}
Two graphs $G,H$ will be considered to be isomorphic ($G \simeq H$) if two bijective mappings exist for vertices $\iota_v : G.V \leftrightarrow H.V$ and edges $\iota_e : G.E \leftrightarrow H.E$ with matching labels, sources and targets, i.e., $\forall v \in G.V : G.\lambda_v(v) = H.\lambda_v(\iota_v(v))$ and $\forall e \in G.E : G.\lambda_e(e) = H.\lambda_e(\iota_e(e)) \wedge G.s(e) = H.s(\iota_e(e)) \wedge G.t(e) = H.t(\iota_e(e))$. 
\end{definition}
%

\begin{definition}
\label{def:lattice}\textsc{(Pattern Lattice).} 
A \textit{pattern} is a connected graph isomorphic to a subgraph $P \simeq S$. Let $\mathcal{P} = \{P^{-1}, P_0,.., P_n\}$ be the set of all patterns isomorphic to any subgraph in a graph collection, than patterns form a \textit{lattice} based on parent-child relationships. $P_p $ will be a parent of $P_c$ if $P_p \sqsubset P_c \wedge |P_p.E| = |P_c.E|$ - 1.
 Based on edge count $k$ there are disjoint \textit{levels} $\mathcal{P}^{-1},.. ,\mathcal{P}^k \subseteq \mathcal{P}$. Root level $\mathcal{P}^{-1} = \{P^{-1}\}$ contains only the empty pattern $P^{-1}$ which is the parent of all patterns with $k=0$. For all other patterns $\forall P^k \in \mathcal{P}, k > 0 \ \exists\ P^{k-1}\in \mathcal{P} : P^{k-1} \sqsubset P^{k}$ is true.
\end{definition}

\hspace{-4mm}
Figure \ref{fig:lattice} shows the lattice of patterns $\mathcal{P} = \{P_{00},..,P_{20}\}$ occurring in the example graph collection $\mathcal{G}$.

\begin{definition}
\label{def:embedding}\textsc{(Embedding).} 
Let $G$ be a graph and $P$ be a pattern, then an \textit{embedding} is defined to be a pair $m(G,P) = \langle \iota_v, \iota_e \rangle$ of isomorphism mappings describing a subgraph $S \sqsubseteq G$ isomorphic to $P$. As a graph may contain $n$ subgraphs isomorphic to the same pattern (e.g., subgraph automorphisms), we use $\mu : \mathcal{P} \rightarrow M^n$ to denote an \textit{embedding map}, which assoicates $n$ elements of an embeddings set $M$ to every pattern $P \in \mathcal{P}$. If $\mu$ maps to an empty tuple, the graph will not contain a pattern.
\end{definition}

\hspace{-4mm}
Figure \ref{fig:lattice} shows three differently colored edge  mappings of example embeddings $m_0(G_1, P_{20}), m_1(G_2, P_{20})$ and $m_2(G_2, P_{20})$.

\begin{definition}
\label{def:support}\textsc{(Frequency/Support).} 
Let $\mathcal{G} = \{G_0,..,G_n\}$ be a graph collection and $P$ be a pattern, then the \textit{frequency} $\phi : \mathcal{P} \rightarrow \mathbb{N}$ of a pattern is the number of graphs containing at least one subgraph isomorphic to the pattern. The term \textit{support} describes the frequency of a pattern relative to
the number of graphs $\sigma(P) = \phi(P) / |\mathcal{G}|$.
\end{definition}

\begin{definition}
\label{def:fsm}\textsc{(Frequent Subgraph Mining).} 
Let $\mathcal{G}$ be a graph collection, $\mathcal{P}$ the set of all contained patterns and $s_{min}$ be the minimum support with $0 \leq s_{min} \leq 1$, then the problem of \textit{frequent subgraph mining} is to identify the complete set of patterns $\mathcal{F} \subseteq \mathcal{P}$ where $\forall P \in \mathcal{P} : P \in \mathcal{F}\Leftrightarrow \sigma(P) \geq s_{min}$ is true.
\end{definition}

\hspace{-4mm}
Frequent subgraph mining for the example graph collection $\mathcal{G} = \{G_1, G_2,G_3\}$ with $s_{min} = 50\% / f_{min} = 2$
results in the five non-empty patterns with $\phi(P) \geq 2$ in the lattice of Figure \ref{fig:lattice}.

\subsection{Related Work}

A recent survey \cite{jiang2013survey} by Jiang et al. provides an extensive overview about frequent subgraph mining (FSM). Due to limited space and the vast amount of work related to this problem we only discuss approaches matching Definition \ref{def:fsm}. Thus, we omit the single-graph setting \cite{bringmann2008frequent, elseidy2014grami, teixeira2015arabesque} as well as graph-transaction approaches with incomplete results like maximal \cite{thomas2010margin}, closed \cite{yan2003closegraph} or significant \cite{ranu2009graphsig} frequent subgraph mining.

The first exact FSM algorithms, e.g., AGM \cite{inokuchi2000apriori} and FSG \cite{kuramochi2001frequent}, followed an \textit{a priori} approach. These algorithms implement a level-wise breath-first-search (BFS, illustrated by Figure \ref{fig:bfs}) in the pattern lattice, i.e., candidate patterns $\mathcal{P}^k$ are generated and the support is calculated by subgraph isomorphism testing. In a subsequent pruning step frequent patterns $\mathcal{F}^k \subseteq \mathcal{P}^k$ are filtered and joined to form children $\mathcal{P}^{k+1}$ (next round's candidates). The search is stopped as soon as $\mathcal{F}^k = \emptyset$. The disadvantage of these algorithms is that they are facing the subgraph isomorphism problem during candidate generation and support counting. Further on, it is possible that many generated candidates might not even appear.

\begin{figure}[t]

\caption{Pattern lattice search strategies.}
\label{fig:searches}
\setcounter {figure} {1} 

\begin{subfigure}[t]{0.15\textwidth}
	\centering
	\caption{BFS}
		\label{fig:bfs}
  \includegraphics[height=2cm]{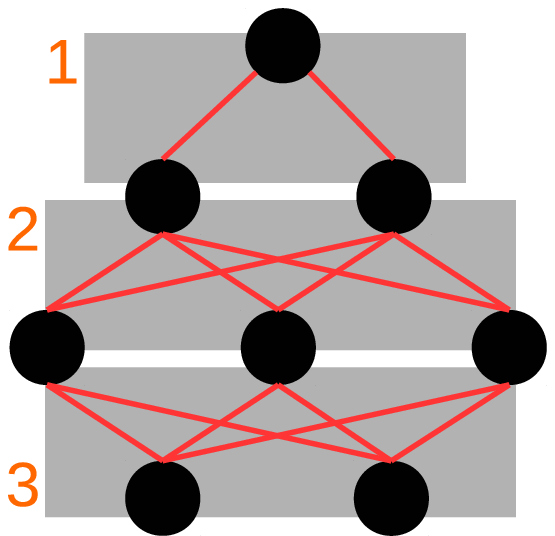}

\end{subfigure}
\begin{subfigure}[t]{0.15\textwidth}
	\centering
	\caption{DFS}
	\label{fig:dfs}
  \includegraphics[height=2cm]{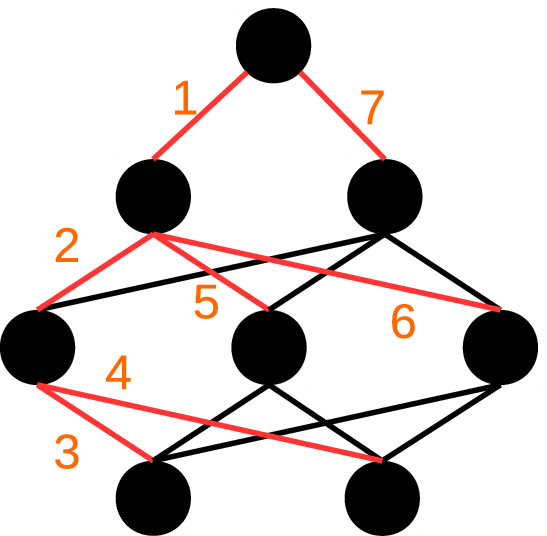}

\end{subfigure}
\begin{subfigure}[t]{0.15\textwidth}
	\centering
	\caption{LDFS}
		\label{fig:ldfs}
  \includegraphics[height=2cm]{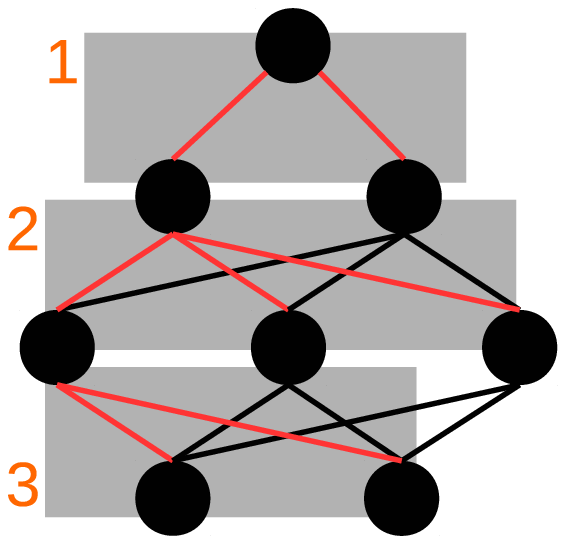}

\end{subfigure}

\end{figure}

Thus, the next generation of \textit{pattern-growth} based FSM algorithms appeared and outperformed the a priori ones. Popular representatives of this category are MOFA \cite{borgelt2002mining}, gSpan \cite{yan2002gspan}, FFSM \cite{huan2003efficient} and Gaston \cite{nijssen2005gaston}. In comparison to the a priori ones, these algorithms traverse the lattice in a depth-first search (DFS, illustrated by Figure \ref{fig:dfs}) and skip certain links in the lattice (dotted lines in Figure \ref{fig:lattice}) to avoid visiting child patterns multiple times. A key concept of these algorithms are canonical labels generated during DFS. However, if labels are generated without recalculation (e.g., gSpan) they won't totally prevent false positives (non canonical labels) and thus an additional isomorphism-based verification will be required. Comparative work \cite{worlein2005quantitative, nijssen2006frequent} has shown that runtime can be decreased by fast label generation and holding embeddings in main memory.

While most popular exact FSM algorithms are from the first half of the 2000s, more recent work focuses on problem variations \cite{jiang2013survey} as well as parallelization, for example, using GPUs \cite{kessl2014parallel}, FPGAs \cite{stratikopoulos2014hpc} and multithreading \cite{vo2015parallel}. All existing approaches of graph transaction FSM on shared nothing clusters \cite{hill2012iterative, lu2013efficiently, aridhi2014novel, lin2014large,  bhuiyan2015iterative} are based on MapReduce \cite{dean2008mapreduce} and will be further discussed in comparison to DIMSpan in Section \ref{sec:mr}. Graph transaction FSM cannot benefit from vertex-centric graph processing approaches \cite{mccune2015thinking} as partitioning a single graph shows different problems than partitioning a graph collection.

\section{Algorithm}
\label{sec:algorithm}

In the following, we provide details about the DIMSpan algorithm including its concept (\ref{sec:concept}), the respective dataflow program (\ref{sec:flow}) as well as pruning and optimization techniques (\ref{sec:branch} - \ref{sec:compression}).

\subsection{Concept}
\label{sec:concept}

In an FSM algorithm that follows the distributed dataflow programming model the input graph collection $\mathcal{G}$ is represented by a dataset of graphs equally partitioned into disjoint subsets $\mathcal{G}_1, \mathcal{G}_2, ..,\mathcal{G}_n$ corresponding to the availble \textit{worker threads} $W = \{w_1,w_2,..,w_n\}$. Thus, transformations can be executed on $\left| W \right|$ graphs in parallel but every exchange of global knowledge (e.g., local pattern frequencies) requires synchronization barriers in the dataflow program which cause network traffic. Our major optimization criteria were minimizing delays dependent on exchanged data volume and, as FSM contains the NP-complete subgraph isomorphism problem, minimize the number of isomorphism resolutions.

To achieve the latter, we adapted approaches of two efficient pattern-growth algorithms gSpan \cite{yan2002gspan} and Gaston \cite{nijssen2005gaston}. These algorithms basically are iterations of pattern growth, counting and filter operations but differ in detail. gSpan allows fast append-only generation of canonical labels representing patterns but records only pattern-graph occurrence lists. This requires subgraph isomorphism testing to recover embeddings. In contrast, Gaston has a more complex label generation tailored to the characteristics of molecular databases but stores complete pattern-embedding maps. For the design of DIMSpan, we combine the strong parts of both algorithms. In particular, we use a derivate of gSpan canonical labels (Section \ref{sec:gspan}) but also store embedding maps to avoid subgraph isomorphism testing at the recovery of previous iterations' embeddings. To minimize the additional memory usage, we use optimized data structures and compression (Section \ref{sec:compression}).

With regard to the absence of shared memory in distributed dataflows, the DFS search of pattern growth algorithms is not optimal as it requires $|\mathcal{P}|$ iterations (one for each visited pattern) while a BFS search only takes $k^{max}$ iterations (maximum edge count). Thus, we decided to perform a \textit{level-wise depth-first search} (LDFS, illustrated by Figure \ref{fig:ldfs}), which can be abstracted as a set of massive parallel constrained DFSs with level-wise forking. This approach allows us to benefit from the efficiency of pattern growth algorithms and to apply level-wise frequency pruning at the same time. For example, in Figure \ref{fig:lattice} we apply the frequency pruning of $P_{10}, P_{11}, P_{12}$ in parallel within the same iteration but use search constraints (Section \ref{sec:branch}) to grow only from $P_{10}$ to $P_{20}$. 

By using a distributed in-memory dataflow system instead of MapReduce, DIMSpan further benefits from the capability to hold graphs including supported patterns and their embeddings in main memory between iterations and to exchange global knowledge by sending complete copies of each iteration's $k$-edge frequent patterns to every worker without disk access. In Apache Spark and Apache Flink this technique is called broadcasting\footnote{http://spark.apache.org/docs/latest/programming-guide.html\#broadcast-variables}\footnote{https://ci.apache.org/projects/flink/flink-docs-release-1.2/dev/batch/index.html \#broadcast-variables}.

\subsection{Distributed Dataflow}
\label{sec:flow}

Algorithm \ref{alg:flow} shows the distributed dataflow of DIMSpan. Inputs are a dataset of graphs $\mathcal{G}$ and the minimum frequency threshold $f_{min}$. The output is the dataset of frequent patterns $\mathcal{F}$. For each graph, supported 1-edge patterns $\mathcal{P}^1$ and the embedding map $\mu^1$ are already computed in a preprocessing step (see Section \ref{sec:dictionary}). Our algorithm is iterative and per iteration one level of the pattern lattice is processed until no more frequent patterns exist (line 12). In the following, we describe transformations and intermediate datasets of the iteration body (lines 4 to 11) in more detail:

\textbf{Line 4 - Report:} In the beginning of each iteration every graph reports all $k$-edge ($k\geq1$) supported patterns, i.e., the keys of the last iteration's embedding map $\mu^k$, through a \textit{flatmap} transformation.

\textbf{Line 5 - Combine:} The partition frequency of patterns $\phi_w : \mathcal{P} \times W \rightarrow \mathbb{N}$ is counted in a \textit{combine} tranformation. As this is the last operation before data is shuffled among workers the execution cardinality of pattern operations (e.g. verification, see Section \ref{sec:validation}) here is already reduced from $\left| \mathcal{P} \times \mathcal{G} \right|$ to $\left| \mathcal{P} \times W \right|$.

\textbf{Line 6 - Reduce:} The global frequency of patterns $\phi : \mathcal{P} \rightarrow \mathbb{N}$ is calculated in a \textit{reduce} transformation. Therefore, partition frequencies are shuffled among workers and summed up. 

\textbf{Line 7 - Frequency pruning:} After global frequencies of all patterns are known, a \textit{filter} transformation is used to determine the frequent ones. Executing  pattern operations here further reduces their cardinality from  $\left| \mathcal{P} \times W \right|$ to  $\left| \mathcal{F} \right| \leq \left| \mathcal{P} \right|$. 

\textbf{Line 8 - Broadcasting:} After $\mathcal{F}^k$ is known, a complete copy is sent to the main memory of all workers using \textit{broadcasting}.

\textbf{Line 9 - Pattern growth:} Here, the previously broadcasted set $\mathcal{F}^k$ is used to filter each graph's embeddings $\mu^k$ to those of frequent patterns. For each of the remaining embeddings, the constrained pattern growth (Section \ref{sec:branch}) is performed to generate $\mu^{k+1}$.

\textbf{Line 10 - Obsolescence filter:} After pattern growth, we apply another \textit{filter} operation and only graphs with non-empty $\mu^{k+1}$ will pass. Thus, $\mathcal{G}$ can potentially shrink in each iteration if only a subset of graphs accumulates frequent patterns. 

\textbf{Line 11 - Result storage:} Finally, we use a binary \textit{union} transformation to add the iteration's results to the final result set.

\begin{algorithm}[t]
\caption{Distributed FSM dataflow.}
\begin{algorithmic}[1]
\REQUIRE{$\mathcal{G} = \{\langle G,\mu^1 \rangle_{i}\}_{i \subset \mathbb{N}}, f_{min}$}
\STATE $\mathcal{F} \leftarrow  \emptyset$
\STATE $\mathcal{F}^k \leftarrow  \emptyset$
\REPEAT
  \STATE $\mathcal{P}^{k} \leftarrow \mathcal{G}.\textbf{flatmap}(report)$
  \STATE $\phi^k_{w} \leftarrow \mathcal{P}^{k}.\textbf{combine}(count)$
  \STATE $\phi^k \leftarrow \phi_w^k.\textbf{reduce}(sum)$
  \STATE $\mathcal{F}^k = \mathcal{P}^{k}.\textbf{filter}(\phi^k(P) \geq f_{min})$
  \STATE \textbf{broadcast}($\mathcal{F}^k$)
  \STATE $\mathcal{G} \leftarrow \mathcal{G}.\textbf{map}(patternGrowth)$
  \STATE $\mathcal{G} \leftarrow \mathcal{G}.\textbf{filter}(\exists\ P : |\mu^{k+1}(G, P)| > 0)$
  \STATE $\mathcal{F} \leftarrow \mathcal{F} \cup \mathcal{F}^k$
\UNTIL{$\mathcal{F}^k \neq \emptyset$}
\RETURN $\mathcal{F}$

\end{algorithmic}
\label{alg:flow}
\end{algorithm}

\subsection{Canonical Labels for Directed Multigraphs}
\label{sec:gspan}

We use a derivate of the gSpan minimum DFS code \cite{yan2002gspan} as canonical labels for directed multigraph patterns:

\begin{definition}
\label{def:dfscode}\textsc{(DFS Code)}.
A \textit{DSF code} representing a pattern of $j$ vertices and $k$ edges ($j,k \geq 1$) is defined to be a $k$-tuple $C = \langle x_1,x_2,..,x_k \rangle$ of extensions, where each \textit{extension} is a hextuple $x =\langle t_a, t_b, l_a, d, l_e, l_b \rangle$ representing the traversal of an edge $e$ with label $l_e \in \mathcal{L}_e$ from a \textit{start} vertex $v_a$ to an \textit{end} vertex $v_b$. $d \in \{in, out\}$ indicates if the edge was traversed in or against its direction. A traversal will be considered to be in direction, if the start vertex is the source vertex, i.e., $v_a(x) = s(e)$. The fields $l_a, l_b \in \mathcal{L}_v$ represent the respective labels of both vertices and their initial discovery times $t_a, t_b \in T \mid T = \langle 0, .., j \rangle$ where the vertex at $t=0$ is always the start vertex of the first extension. A DFS code $C_p$ will be considered to be the parent of a DFS code $C_c$, iff $\forall i \in \langle 1,..,k-1 \rangle : C_c.x_i = C_p.x_i$.
\end{definition}

According to this definition child DFS codes can be easily generated by adding a single traversal to their parent. Further on, DFS codes support multigraphs since extension indexes can be mapped to edges identifiers to describe embeddings.

However, there may exist multiple DFS codes representing the same graph pattern. To use DFS codes as a canonical form, gSpan is using a lexicographic order to determine a minimum one among all possible DFS codes \cite{yan2002tr}. This order is a combination of two linear orders. The first is defined on start and end vertex times of extensions $T \times T$, for example, a backwards growth to an already discovered vertex is smaller than a forwards growth to a new one. 
The second order is defined on the labels of start vertex, edge and end vertex $\mathcal{L}_v \times \mathcal{L}_e \times \mathcal{L}_v$, i.e., if a comparison cannot be made based on vertex discovery times, labels and their natural order (e.g., alphabetical) are compared from left to right. To support directed graphs, we extended this order by direction $D = \{in, out\}$  with $out < in$ resulting into an order over $\mathcal{L}_v \times D \times \mathcal{L}_e \times \mathcal{L}_v$, i.e., in the case of two traversals with same start vertex labels, a traversal of an outgoing edge will always be considered to be smaller.

\begin{definition}
\label{def:mindfs}\textsc{(Minimum DFS Code)}.
There exists an order among DFS codes such that $\forall C_1, C_2 : C_1 < C_2 \vee C_1 = C_2 \vee C_1 > C_2$ is true. Let $\mathcal{C}_P$ be the set of all DFS codes describing a pattern P and $C_{min}$ be its minimum DFS code, than $\nexists\ C_i \in \mathcal{C}_P : C_i < C_{min}$ is true.
\end{definition}


%
%

%

\begin{figure}[t]
\vspace{-3mm}
\begin{algorithm}[H]
\begin{algorithmic}[1]
\REQUIRE{$G, \mu^k, \mathcal{F}^k = \langle P_0,..,P_n \mid \text{sorted by } C_{min}\rangle$}
\STATE $C^1_{min} \leftarrow \langle \rangle$ \ \ \ \ \ \ \ \ \ // minimum branch
\STATE $E_{\geq min} \leftarrow G.E$ \ \ \ \ // shrinking branch-validated edge set
\FOR{$P^k \in \mathcal{F}^k \mid \mu^k(G, P^k) \neq \langle\rangle$}
  \IF{$C^1(P^k) > C^1_{min}$}
    \STATE $C^1_{min} \leftarrow C^1(P^k)$\ \ // update min branch and edge set
    \STATE $E_{\geq min} \leftarrow \subset E_{\geq min} \mid C^1(e) \geq C^1_{min}$
  \ENDIF
  \FOR{$m^k, e \in (\mu^k(G, P^k) \times E_{\geq min})$}
    \IF{$\nexists\ m^k.\iota_e(e)$ \textbf{and} time constraint satisfied}
    \STATE grow $P^{k+1}, m^{k+1}$ and add to $\mu^{k+1}$
    \ENDIF
  \ENDFOR
\ENDFOR
\RETURN $G, \mu^{k+1}$
\end{algorithmic}
\caption{Pattern growth map function $\tau$.}
\label{alg:pg}
\end{algorithm}
\vspace{-5mm}
\end{figure}

\subsection{Constrained Pattern Growth}
\label{sec:branch}

Besides gSpan's canonical labels we also adapted the growth constraints to skip parent-child relationships in the pattern lattice (dotted lines in Figure \ref{fig:lattice}). However, in contrast to gSpan, we don't perform a pattern-centric DFS (Figure \ref{fig:dfs}) but an level-wise DFS (Figure \ref{fig:ldfs}), i.e., we perform highly concurrent embedding-centric searches. Due to limited space, we refer to \cite{yan2002tr} for the theoretical background and focus on our adaptation to the distributed dataflow programming model.

There are two constraints for growing children of a parent embedding. The first, in the following denoted by \textit{time constraint}, dictates that forwards growth is only allowed starting from the rightmost path and backwards growth only from the rightmost vertex, where \textit{forwards} means an extension to a vertex not contained in the parent, \textit{backwards} means an extension to a contained one, the \textit{rightmost vertex} is the parent's latest discovered vertex and the \textit{rightmost path} is the path of forward growths from the initial start vertex to the rightmost one. The second constrained, in the following denoted by \textit{branch constraint}, commands that the minimum DFS code of an edge $C^1(e)$ needs to be greater than or equal to the parent's \textit{branch} $C^1(P)$ which is the 1-edge code described by only the initial extension of the a pattern's minimum DFS code.

Algorithm \ref{alg:pg} shows our adaption of these constraints to the distributed dataflow programming model, in particular, a map function $\tau$ that executes pattern growth for all embeddings of frequent patters in a single graph (line 9 of Algorithm \ref{alg:flow}). Therefore, we hold not only $G$ but also embedding map $\mu^k$ for each element of $\mathcal{G}$ and enable $\tau$ access to $\mathcal{F}^k$ as it was received by every worker in the broadcasting step (line 8 of Algorithm \ref{alg:flow}). 

In an embedding-centric approach, a naive solution would be testing possible growth for the cross of supported frequent patterns' embeddings and the graph's edges. As an optimization, we use a merge strategy based on the branch constraint to reduce the number of these tests. Therefore, $\mathcal{F}^k$ in Algorithm \ref{alg:pg} is an n-tuple and ascendantly sorted by minimum DFS code. When executing the map function, we keep a current minimum branch $C^1_{min}$ and a current edge candidate set $E_{\geq min}$ (lines 1,2). Then, for every supported frequent pattern (line 3) we compare its branch to the current minimum (line 4) and only if it is greater, the current minimum will be updated (line 5) and the set of growth candidates can be shrunk (line 6). Thus, only for the cross of embeddings and branch-validated edges (line 8) parent containment and the time constraint need to be checked (line 9). In the case of a successful growth (line 10) the resulting pattern and its embedding will be added to $\mu^{k+1}$, the output of the map function (line 14). Sorting and rightmost path calculation are not part of the map function and executed only $|W \times \mathcal{F}|$ times at broadcast reception.

\vspace{-2mm}
\subsection{False Positive Verification}
\label{sec:validation}
\vspace{-2mm}
Although the constrained pattern growth described previously helps skipping links in the pattern lattice (dotted lines in Figure \ref{fig:lattice}), it gives no guarantee for visiting every pattern only once. In the case of multiple ($n$) visits, $n-1$ non-minimal DFS codes (\textit{false positives}) will be generated. Thus, they need to be verified, e.g., by turning the label into a graph and recalculating the minimum DFS code. This is the only part of the algorithm facing the isomorphism problem and reducing its cardinality may reduce total runtime \cite{yan2002tr}. Thus, we evaluated moving the verification step to three different steps in the dataflow, in particular before reporting (line 4 in Algorithm \ref{alg:flow}), after partition frequency counting (line 5) and after frequency pruning (line 7). In the first case, false positives won't be counted and shuffled but verification is executed $|\mathcal{P} \times \mathcal{G}|$ times; in the second case, false positives are counted but not shuffled with $| \mathcal{P} \times W|$ verifications and in the last case, they will be counted and shuffled but only $|\mathcal{F}|$ verifications are required. By experimental evaluation we found that the first option in always slow while the others lead to similar runtimes (see Section \ref{sec:config}).

\vspace{-2mm}
\subsection{Preprocessing and Dictionary Coding}
\label{sec:dictionary}
\vspace{-1mm}
Before executing the dataflow shown by Algorithm \ref{alg:flow}, we apply preprocessing that includes label-frequency based pruning, string-integer dictionary coding and  sorting edges according to their 1-edge minimum DFS codes. The original gSpan algorithm already used these concepts but we improved the first two and adapted the third to our level-wise DFS strategy. In the first preprocessing step, we determine frequent vertex labels and broadcast a dictionary to all workers. Afterwards, we drop all vertices with infrequent labels as well as their incident edges. Then, we determine frequent edge labels, in contrast to the original, only based on the remaining edges. Thus, we can potentially drop more edges, for example, $e_1$ of $G_1$ in Figure \ref{fig:lattice} would be removed. This would not be the case by just evaluating its edge label since without dropping $e_2$ of $G_0$ before (because $v_2$ has infrequent label \texttt{C}) the frequency of edge label \texttt{b} would be $2$, i.e., considered to be frequent.

After dictionaries for vertex and edge labels are made available to all workers by broadcasting, we not only replace string labels by integers to save memory and to accelerate comparison but also sort edges according to their minimum DFS code, i.e., we use n-tuples instead of sets to store edges. We benefit from the resulting sortedness in every execution of the constrained pattern growth (see Section \ref{sec:branch}) as the effort of determining branch-valid edge candidates (line 6 of Algorithm \ref{alg:pg}) is reduced from a set filter operation to a simple increase of the minimum edge index.

\begin{figure}[t]
  \caption{Dataset element representing graph $G_2$, pattern $P_{20}$ and embedding set $M(G_2, P_{20})$ of Figure \ref{fig:lattice}.}
	\label{fig:element}
	\centering
  \includegraphics[width=0.48\textwidth]{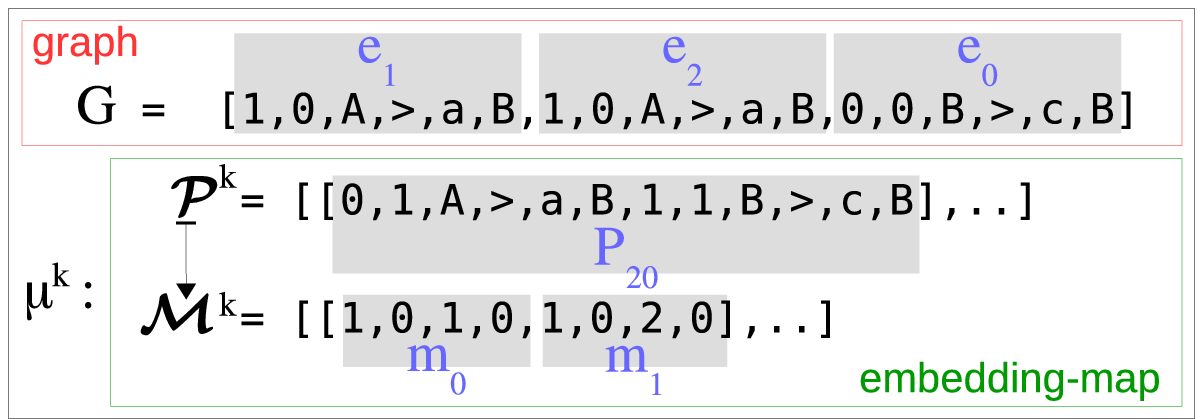}

\end{figure}

\subsection{Data Structures and Compression}

\label{sec:compression}
We not only use minimum DFS codes as canonical labels but also a data structure based thereon to support all pattern operations (counting, growth and verification) without format conversions. We further store graphs as sorted lists of 1-edge DFS codes to allow a direct comparison at the lookup for the first valid edge of a branch in the pattern growth process (line 6 of Algorithm \ref{alg:pg}). Figure \ref{fig:element} illustrates a single element of $\mathcal{G}$ in Algorithm \ref{alg:flow} representing $G_2$ from Figure \ref{fig:lattice} and its embedding map $\mu^k$ in the $k=2$ iteration. Graphs and patterns are stored according to Definition \ref{def:dfscode} but encoded in integer arrays where all 6 elements store a graph's edge or a pattern's extension. For the sake of readability we use alphanumerical characters in Figure \ref{fig:element}. $\mu^k$ is stored as a pair of nested integer arrays $\langle \mathcal{P}^k, \mathcal{M}^k \rangle$ where equal indexes map embeddings to patterns. All embeddings of the same pattern are encoded in a single multiplexed integer array where all $|P.V|+|P.E|$ elements store a single embedding. Here, indexes relative to their offset relate vertex ids to their initial discovery time and edge ids to extension numbers. 

This data structure not only allows fast pattern operations but also enables lightweight and effective integer compression. Therefore, we exploit the predictable value ranges of our integer arrays. As we use dictionary coding and vertex discovery times are bound by the maximum edge count $k_{max}$ the array's values may only range from $0..(max(k_{max},l_v, l_e)-1)$ where $l_v, l_e$ are the numbers of distinct vertex and edge labels. In the context of FSM, the maximum value will typically be much less than the integer range of $2^{32}$. There are compression techniques benefiting from low-valued integer arrays \cite{lemire2015decoding}. In preliminary experiments we found that Simple16 \cite{zhang2008performance} allows very fast compression and gives an average compression ratio of about 7 over all patterns found in our synthetic test dataset (see Section \ref{sec:data}). We apply integer compression not only to patterns but also to graphs and embeddings, which also have low maximum values, to decrease memory usage. Embeddings and graphs are only decompressed on demand and at maximum for one graph at the same time. All equality-based operations (map access and frequency counting) are performed on compressed values. Our experimental evaluation results show a significant impact of this compression strategy (see Section \ref{sec:config}).

\begin{table*}[t]
\centering
\caption{Comparison to MapReduce approaches by upper bounds of disk access (teletype font), network traffic (italic font) and isomorphism resolution cardinality (bold font).}
\label{tab:comp}

\begin{tabular}{|c|lll|l|ll}
\hline
&
\multicolumn{3}{|c|}{\textit{iterative}}
& \textit{1-phase}
\\
& I-FSM \cite{hill2012iterative} 
& MR-FSE \cite{lu2013efficiently} 
& DIMSpan
& F\&R \cite{lin2014large} 
\\
\hline
\hline
Pre
& 
& 
& \texttt{read} $\mathcal{G}$ 
& 
\\
\hline
\hline
Map
& \texttt{read} $\mathcal{S}$
& \texttt{read} $\mathcal{M}$, $W \times \mathcal{P}$
& 
& \texttt{read} $\mathcal{G}$
\\
1
& 
& \textbf{growth} $|\mathcal{S}|$ 
& 
& \textbf{Gaston} $|W \times \mathcal{P}|$ 
\\
& 
& \texttt{write} $\mathcal{M}$
& 
& 
\\
\hline
Red.
& \textit{shuffle} $\mathcal{S}$
& 
&
& \textit{shuffle} $W \times \mathcal{P}$
\\
1
& \texttt{write} $\mathcal{S}$
& 
&
& \texttt{write} $ \mathcal{P}$
\\
\hline
\hline
Map
& \texttt{read} $\mathcal{S}$
& \texttt{read} $\mathcal{M}$
&
& \texttt{read} $\mathcal{G}$,  $W \times  \mathcal{P}$
\\
2
& \textbf{label} $|\mathcal{S}|$
& 
& \textbf{verify} $|W \times \mathcal{P}|$
& \textbf{ref} $(|\mathcal{G}| - 1) * |\mathcal{P}|$
\\
\hline
Red.
& \textit{shuffle} $\mathcal{S}$
& \textit{shuffle} $\mathcal{I}_\mathcal{G} \times \mathcal{P}$
& \textit{shuffle} $W \times \mathcal{P}$
& \textit{shuffle} $W \times \mathcal{P}$
\\
2
& \texttt{write} $\mathcal{S}$
& \texttt{write} $\mathcal{P}$
& \textit{send} $W \times \mathcal{P}$
& \texttt{write} $\mathcal{P}$ 
\\
\hline
\hline
Post
& 
& 
& \texttt{write} $\mathcal{P}$ 
& 
\\
\hline
\end{tabular}

\hspace{-4cm}
\begin{tabular}{ll}
data volume: & $\mathcal{S} > \mathcal{M} > (\mathcal{I}_\mathcal{G} \times \mathcal{P}) \gg (W \times \mathcal{P}) >  \mathcal{P}$\\
cardinality: & $|\mathcal{S}| > ((|\mathcal{G}| - 1) * |\mathcal{P}|) \gg |W \times \mathcal{P}|$  \\
\end{tabular}
\end{table*}

\section{Comparison to Approaches based on MapReduce}
\label{sec:mr}

To the best of our knowledge, only five approaches to transactional FSM based on shared nothing clusters exist \cite{hill2012iterative, lu2013efficiently, aridhi2014novel, lin2014large, bhuiyan2015iterative}. They are all based on MapReduce. In this section, we compare three of these approaches to DIMSpan since \cite{aridhi2014novel, bhuiyan2015iterative} show relaxed problem definitions in comparison to Definition \ref{def:fsm}. The comparison focuses on our optimization criteria, in particular upper bounds of shuffled data volume, required disk access and the number of isomorphism resolutions. Isomorphisms are resolved either when counting patterns by subgraph isomorphism testing or, as both require enumerating all permutations of a certain graph representation, at the generation of canonical labels from scratch as well as during their verification (see Section \ref{sec:validation}) . 

\subsection{Comparison}

Table \ref{tab:comp} shows a comparison of I-FSM \cite{hill2012iterative}, MR-FSE \cite{lu2013efficiently}, DIMSpan and the filter-refinement (F\&R) approach of \cite{lin2014large} with regard to the stated dimensions. While the first three are iterative (i.e., level-wise search), F\&R is partition-based and requires only a single phase. All approaches including DIMSpan can be represented by two map-reduce (MRMap-reduce) phases where upper bounds of iterative approaches express the union of single iterations. On top of Table \ref{tab:comp} we provide orders among data volumes and cardinalities.

\textbf{I-FSM} is using complete subgraphs $\mathcal{S}$ as its main data structure. In Map 1 $k$-edge subgraphs of the previous iteration are read from disk and shuffled by graph id. In Reduce 1, graphs are reconstructed by a union of all subgraphs. Afterwards, $k+1$-edge subgraphs are generated and written to disk. In Map 2 they are read again and a (in \cite{hill2012iterative} not further specified) canonical label is calculated for every subgraph. Thus, the isomorphism problem is resolved with maximum cardinality $|\mathcal{S}|$. Then, all subgraphs are shuffled again according to the added label. In Reduce 2, label frequencies are counted. Finally, all subgraphs showing a frequent label are written to disk.

\textbf{MR-FSE} is using embedding maps $\mathcal{M}$ as its main data structure, i.e., with regard to vertex- and edge labels an irredundant version of $\mathcal{S}$ that describes subgraphs by patterns and embeddings (see Section \ref{sec:problem}). In Map 1 $k$-edge maps of the previous iteration are read from disk. Additionally, all $k$-edge frequent patterns are read by each worker ($W \times \mathcal{P}$). Then, graphs are reconstructed based on embeddings, pattern growth is applied and updated maps are written back to disk. MR-FSE is using DFS codes like DIMSpan but in \cite{lu2013efficiently} it is clearly stated that no verification is performed at any time. Instead, false positives are detected by enumerating all DFS code permutations of each distinct edge set (subgraph) to choose the minimal one. Consequently, isomorphisms among DFS codes are resolved $|\mathcal{S}|$ times. Reduce 1 is not used. In Map 2 the grown maps are read again and a tuple for each pattern and supporting graph ($\mathcal{I}_\mathcal{G} \times \mathcal{P})$ is shuffled. In Reduce 2, pattern frequencies are counted, filtered and written to disk.
 
\textbf{F\&R} reads graphs from disk and runs Gaston \cite{nijssen2005gaston}, an efficient single-machine algorithm, on each partition in Map 1. Then, a statistical model is used to report partition frequencies of patterns. Thus, every pattern is verified and shuffled only once per partition ($W \times \mathcal{P}$). In Reduce 1, local frequencies are evaluated for each pattern and a set of candidate patterns $\mathcal{P}$ including some frequency information are written to disk. In Map 2 graphs and information about candidate patterns are read from disk. For some partitions, local pattern frequencies may be unknown at this stage. Thus, they are refined by a priori like subgraph-isomorphism testing. The upper bound is not fully $|\mathcal{G} \times \mathcal{P}|$ as it is guaranteed that the exact frequency is known for at least one partition. In Reduce 2, refined pattern frequencies are summed up, filtered and written to disk.

\begin{table*}[t]

\caption{Runtimes for increasing input and result size.}
\label{tab:size}
\centering


\begin{tabular}{|c|cccc|cccc|}
\hline
\textbf{data} & \multicolumn{4}{c|}{\textbf{molecular}} & \multicolumn{4}{c|}{\textbf{synthetic}} \\
\hline
$s_{min}$ 	&	\textit{30\%} 	&	\textit{10\%} 	&	\textit{5\%} 	&	\textit{3\%} 	&	\textit{100\%} 	&	\textit{90\%} 	&	\textit{70\%} 	&	\textit{30\%} 	\\
$|\mathcal{F}|$			&		\textit{127}		&		\textit{1270}		&		\textit{4660}		&		\textit{12805}		&	 \textit{702} 	&		\textit{1404}		&		\textit{2808}		&		\textit{5616}		\\
\hline
$|\mathcal{G}|$ & \multicolumn{4}{c|}{runtime in minutes} & \multicolumn{4}{c|}{runtime in minutes} \\
\hline
\textbf{100K}&0,6&1,1&2,1&	5,3		&		1,1		      &		1,5		&		2,4		&		4,2		\\
\textbf{333K}		&		0,9		&		2,0		&		4,7		&		15,9	&		2,0		      &		3,2		&		6,3		&		12,2		\\
\textbf{1M}		  &		1,9		&		4,9		&		13,2	&		46,2	&		4,4		      &		7,9		&		17,3		&		34,2		\\
\textbf{3.3M}		&		5,7		&		15,2	&		43,4	&		150,7	&		13,9		    &		24,9		&		52,6		&		103,7		\\
\textbf{10M}	  &	15,4	  &	43,7	  &	125,5	  &	443,6	  &	38,1	&	71,2	&	148,6	&	295,3	\\
\hline
\end{tabular}

\end{table*}

\begin{figure*}[t]
\caption{Relative runtimes per 100K graphs.}
\label{fig:size}
\centering

\begin{subfigure}[t]{0.35\textwidth}
  \includegraphics[height=5cm]{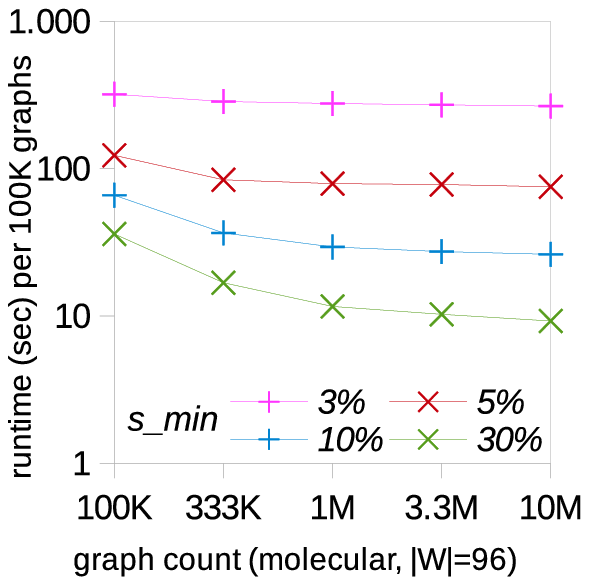}
	\label{fig:s_runtime}
\end{subfigure}
\begin{subfigure}[t]{0.5\textwidth}
  \includegraphics[height=5cm]{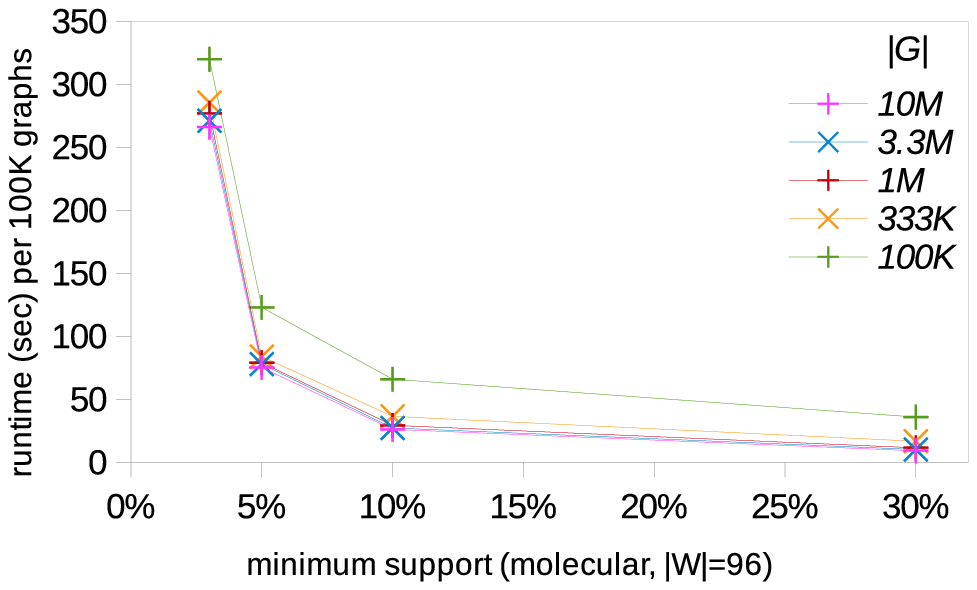}
	\label{fig:s_speedup}
\end{subfigure}
\\

\begin{subfigure}[t]{0.35\textwidth}
  \includegraphics[height=5cm]{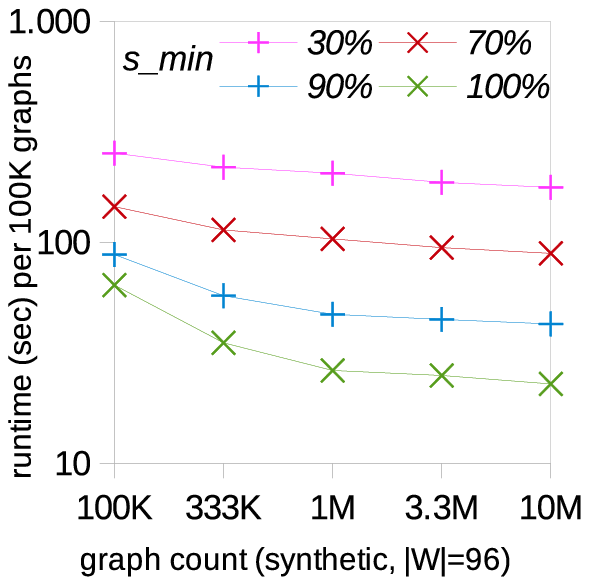}
	\label{fig:s_runtime}
\end{subfigure}
\begin{subfigure}[t]{0.5\textwidth}
  \includegraphics[height=5cm]{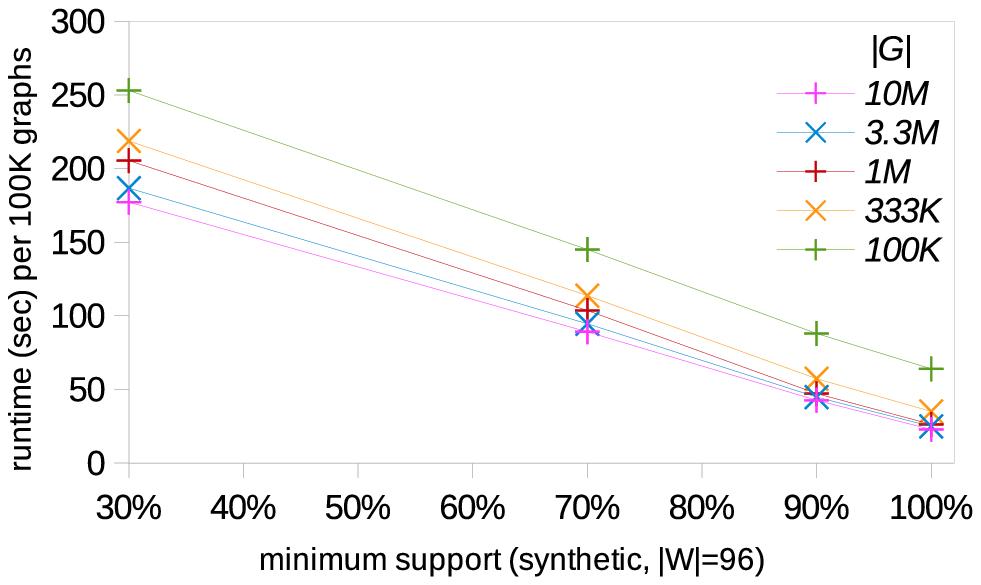}
	\label{fig:s_speedup}
\end{subfigure}

\end{figure*}

\textbf{DIMSpan} reads graphs only once from disk before the iterative part and writes patterns only once to disk afterwards. Map and Reduce 1 are similar to MR-FSE, but DIMSpan keeps graphs as well as embedding maps in main memory and requires no disk access to keep state. In its default configuration, DIMSpan verifies patterns after a combine operation at the end of Map 2 and is thus facing isomorphism resolution only $|W \times \mathcal{P}|$ times. By configuration, verification can be moved to Reduce 2 which further reduces cardinality to $|\mathcal{P}|$ (see Section \ref{sec:validation}). The second effect of the combine operation is shuffling only partition frequencies like F\&R. To make frequent $k$-edge patterns available to all workers, DIMSpan uses broadcasting which requires only network traffic but no disk access.

\subsection{Summary}

DIMSpan reduces disk access to a minimum as it is based on a distributed in-memory system. However, beyond the technology-based advantage, DIMSpan is also superior in the number of isomorphism resolutions as it moves pattern verification after counting and will not apply a priori like operations at any time. Further on, DIMSpan shuffles partition frequencies only once which is less than F\&R (twice) and much less than complete subgraphs (I-FSM) or graph-pattern supports (MR-FSE). Besides the discussed dimensions, only F\&R and DIMSpan use compression. DIMSpan is the only approach that supports directed multigraphs and applies preprocessing (see Section \ref{sec:dictionary}). 

Our comparison clearly shows that I-FSM is inefficient, as it is the only approach that reads, shuffles and writes full subgraphs twice per iteration. On the other hand, it would have been interesting to reproduce evaluation results of MR-FSE and F\&R on our own cluster. Unfortunately, MR-FSE is not available to the public. Regarding F\&R, only binaries\footnote{\url{https://sourceforge.net/projects/mrfsm/}} are available. However, there is no sufficient English documentation and they rely on an outdated non-standard Hadoop installation. Thus, we were not able to execute the binaries without errors despite notable effort and support of the author.

\section{Evaluation}
\label{sec:eval}

In this section we present the results of a performance evaluation of DIMSpan based on a real molecular dataset of simple undirected graphs and a synthetic dataset of directed multigraphs. We evaluate scalability for increasing volume of input, decreasing minimum support and variable cluster size. Further on, we analyze the runtime impact of the discussed pruning and optimization techniques.

\subsection{Implementation and Setup}
\label{sec:impl}

We evaluated DIMSpan using Java 1.8.0\_102, Apache Flink 1.1.2 and Hadoop 2.6.0. More precisely we used Flink's DataSet API\footnote{\url{https://ci.apache.org/projects/flink/flink-docs-release-1.2/dev/batch/index.html}} for all transformations and its \textit{bulk iteration} for the iterative part. We further used the Simple16 implementation from JavaFastPFOR\footnote{\url{https://github.com/lemire/JavaFastPFOR}} for compression. The source code is available on GitHub\footnote{\url{https://github.com/dbs-leipzig/gradoop}; org.gradoop.examples.dimspan} under GPL v3 licence. All experiments were performed on our in-house cluster with 16 physical machines equipped with an Intel E5-2430 2.5 Ghz 6-core CPU, 48 GB RAM, two 4 TB SATA disks and running openSUSE 13.2. The machines are connected via 1 Gigabit Ethernet. 

\begin{figure*}[t]
\centering
\caption{Speedup (su) for varying cluster size.}
\label{fig:speedup}
\begin{subfigure}[t]{0.5\textwidth}
\vspace{-4.5cm}

\begin{tabular}{|l|cc|cc|}
\hline
\textit{data}\ \ \	$|\mathcal{G}|$	&	\textit{mol.}	&	\textit{1M}	&	\textit{syn.}	&	\textit{1M}	\\
$s_{min}$\ \	$|\mathcal{F}|$	&	\textit{5\%}	&	\textit{4660}	&	\textit{70\%}	&	\textit{2808}	\\
\hline
\textbf{size($|W|$)}	&	\textbf{min}	&	\textbf{su}*	&	\textbf{min}	&	\textbf{su}*	\\
\hline
1	(6)	&	131	&		&	254	&		\\
2	(12)	&	71	&	1.0	&	108	&	1.0	\\
4	(24)	&	39	&	1.8	&	56	&	1.9	\\
8	(48)	&	22	&	3.1	&	32	&	3.4	\\
16 (96)	&	13	&	5.5	&	17	&	6.4	\\
\hline
\end{tabular}
\vspace{1mm}
\begin{small}

*over 2 as Flink's execution differs for size 1
\end{small}

\end{subfigure}
\begin{subfigure}[t]{0.3\textwidth}
  \includegraphics[height=5cm]{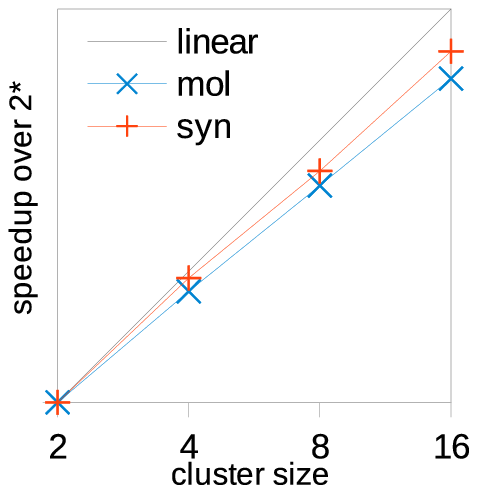}
	\label{fig:s_runtime2}
\end{subfigure}

\end{figure*}

\subsection{Data Sets}
\label{sec:data}

We evaluated three data-related dimensions that impact the runtime of a distributed FSM algorithm: structural characteristics, \textit{input size} $|\mathcal{G}|$ and \textit{result size} $|\mathcal{F}|$. To show scalability in relation to one of the dimensions, the other two need to be fixed. While $|\mathcal{F}|$ can be increased  by decreasing the minimum support threshold, varying the other two dimensions separately is less trivial. Thus, we decided to use two base datasets with divergent structural characteristics and just copy every graph several times to increase $|\mathcal{G}|$ under preservation of structural characteristics and $|\mathcal{F}|$ . 

The first base dataset is \textit{yeast-active}\footnote{\url{https://www.cs.ucsb.edu/~xyan/dataset.htm}}, in the following denoted by \textit{molecular}, a real dataset from anti-cancer research. It was chosen to represent molecular databases because structural characteristics among them do not fundamentally differ due to the rules of chemistry. For example, all molecular databases describe simple undirected graphs with only two different edge labels (single and double bond) and most frequent patterns are paths or trees \cite{nijssen2005gaston}. The base dataset contains around 10K graphs (9567) and is scaled up to datasets containing around 100K to 10M graphs. We did not use an optimized version of DIMSpan for undirected graphs but provide an according parameter. If the parameter is set to undirected, the direction indicator (see Section \ref{sec:gspan}) will just be ignored. Dedicated application logic is only used when it is unavoidable, for example, an 1-edge DFS code desribing a non-loop edge with two equal vertex labels (automorphism) leads to two embeddings in undirected mode.

The second category of datasets, in the following denoted by \textit{synthetic}, was created by our own data generator\footnote{org.gradoop.flink.datagen.transactions.predictable}.  It generates unequally sized connected directed multigraphs where each 10th graph has a different size ranging from $|V| = 10, |E| = 14$ to $|V| = 91, |E| = 140$. There are 11 distinct vertex and $5 + |\mathcal{G}|/1000$ distinct edge labels. The result is predictable and contains 702 frequent patterns with 1 to 13 edges for each min support decrement of 10\% (i.e., 702 for 100\%, 1404 for 90\% , ..). The patterns contain loops, parallel edges (in and against direction), different subgraph automorphisms (e.g., "rotated" and "mirrored") separately as well as in all combinations. The data generator was not only designed for benchmarking but also for testing the correctness of implementations. To verify the number of contained frequent patterns we implemented a simple pruning-free brute-force FSM algorithm and manually verified all patterns of sizes 1..4, 12,13.

\begin{table*}
\begin{center}
\caption{Slowdown (sd) by configuration changes ($|W| = 96$).}
\label{tab:slowdown}

\begin{tabular}{|l|cc|cc|}
\hline
\multicolumn{1}{|r|}{\textit{data} \ \ \ \ \ $|\mathcal{G}|$} &	\textit{mol.}	&	\textit{1M}	&	\textit{syn.}	&	\textit{1M}	\\
\multicolumn{1}{|r|}{$s_{min}$ \ \ \  $|\mathcal{F}|$}	&	\textit{3\%}	&	\textit{12805}	&	\textit{30\%}	&	\textit{5616}	\\
\hline
\textbf{configuration change}& \textbf{min} & \textbf{sd} & \textbf{min} & \textbf{sd} \\

\ \ \ default (with all optimizations) &	45	&		&	34	&		\\
\hline
(1) no pattern compression	  &	64	&	41\%	&	48	&	41\%	\\
(2) pattern compression before shuffle	&	62	&	36\%	&	49	&	44\%	\\
(3) no embedding compression	      &	59	&	30\%	&	37	&	8\%	\\
(4) no graph compression	    &	47	&	4\%	&	34	&	0\%	\\

(5) verification before reporting	    &	54	&	20\%	&	39	&	16\%	\\
(6) verification after frequency filter	  &	46	&	1\%	&	34	&	-2\%	\\

(7) no preprocessing	        &	51	&	12\%	&	73	&	115\%	\\
(8) branch constraint in verification      &	48	&	7\%	&	31	&	-10\%	\\
\hline
\end{tabular}

\end{center}

\end{table*}

\subsection{Input and Result Size}
\label{sec:size}
\vspace{-2mm}
Table \ref{tab:size} and Figure \ref{fig:size} show measurement results for increasing input $|\mathcal{G}|$ and result size $|\mathcal{F}|$ (decreasing minimum support $s_{min}$) for both datasets on a cluster with 16 machines, i.e., 96 worker threads ($|W| = 96$). Table \ref{tab:size} shows absolute runtimes in minutes while Figure \ref{fig:size} illustrates relative runtimes for processing a portion of 100K graphs in seconds, i.e., $t(100K) =  t(|\mathcal{G}|)  * \nicefrac{100K}{|\mathcal{G}|}$. 
In more detail, Figure \ref{fig:size} shows runtimes for an increasing number of graphs (left hand side) and for increasing result sizes (right hand side) for molecular as well as synthetic datasets. We observe that DIMSpan achieves an excellent scalability with regard to both dimensions since for nearly all configurations runtime grows less than the input size or result size.  

The charts on the left hand side of Figure \ref{fig:size} indicate that the time to process 100K graphs is constantly decreasing with an increasing input size for both workloads. The reason is our optimization strategy that verifies DFS codes after counting (see Section \ref{sec:validation}) which makes the number of isomorphism resolutions independent from the input size. However, we see a decrease of this effect with decreasing threshold, which indicates that pattern growth becomes more time consuming for lower thresholds.

The charts on the right hand side of Figure \ref{fig:size} show the time to process 100K graphs for decreasing support thresholds, i.e., increasing result sizes. The shapes of both charts fundamentally differ as the result size of the molecular dataset increases exponentially for decreasing thresholds while the synthetic dataset (by design) shows a linear growth. 
For the synthetic datasets we observe near-perfect linearly increasing runtimes. On the other hand, for our real-world molecular datasets there are non-linear effects for low support thresholds. While runtime grows less than the result size for up to a minimal support $s_{min}$ of 5\%, a further reduction of $s_{min}$ causes a  higher increase in runtimes. For example, for 10M graphs the runtime goes up by factor $3.5$ for $s_{min}=3\%$ compared to 
$s_{min}=5\%$ while the result size only increases by factor $2.7$. Additionally, we see again the positive effect of post-counting verification as scalability becomes better with increased input volume.



\vspace{-2mm}
\subsection{Cluster Size}
\label{sec:scalability}
\vspace{-2mm}
Figure \ref{fig:speedup} shows measured runtimes and gained speedup for varying cluster sizes with fixed $|\mathcal{G}|$ and $s_{min}$. The speedup is measured over cluster size 2, as Flink is choosing an alternative execution strategy for a single machine which would lead to a superlinear speedup from 1 to 2 machines on the synthetic dataset. 
We see that DIMSpan scales sublinear but achieves notable speedups on both datasets for an increasing number of machines which justifies adding machines to decrease absolute runtime in big data scenarios.

\vspace{-1mm}
\subsection{Configuration Slowdown}
\label{sec:config}
\vspace{-1mm}
To analyze the impact of the proposed optimizations, we evaluated to which degree response times slow down when we omit single optimization and pruning techniques.
Table \ref{tab:slowdown} shows the observed slowdowns compared to the default DIMSpan algorithm including all optimizations for 16 machines and fixed data parameters. We see 
that there are some differences between the two datasets, but that pattern compression (1, 2) is the most effective optimization technique. Its effectiveness  is
primarily because of faster counting that is enabled by smaller data objects rather than the lower network traffic. As our integer array representation is already memory efficient, performing compression dedicatedly before shuffling even lead to a larger slowdown for the synthetic dataset as the compression effort is higher than its benefit. Embedding compression (3) and graph compression (4) not only lower the memory footprint but also increase runtime as data passed among iterations can be faster serialized by Apache Flink.

Moving the verification to the end of the pattern growth process (5) will show, as expected, a notable slowdown, even if false positives are counted otherwise (see Sections \ref{sec:validation} and \ref{sec:mr}). Moving the verification after the filter step (6) has no notable impact. The effect of our preprocessing (7) highly depends on the dataset. It is very high for our synthetic dataset with many infrequent edge labels but still effective on the real molecular data. Finally (8), we disabled the branch constraint check (line 4 of Algorithm \ref{alg:pg}). Note, that the result remains correct as the branch pruning also applies automatically at verification. For the synthetic dataset, this leads even to an improved runtime as we avoid sorting edges in the preprocessing, don't perform the check in pattern growth and consequently never execute lines 5 and 6 of Algorithm \ref{alg:pg}. By contrast, we can benefit from this technique for the molecular dataset as it has a much lower number of distinct minimum 1-edge DFS codes.

\vspace{-2mm}
\section{Conclusions \& Future Work}
\label{sec:conclusion}
\vspace{-1mm}
We proposed DIMSpan, the first approach to parallel transactional FSM that combines the effective search space pruning of a leading single-machine algorithm with the technical advantages of state-of-the-art distributed in-memory dataflow systems. DIMSpan is part of \textsc{Gradoop} \cite{junghanns2016epgm, petermann2016graph}, an open-source framework for distributed graph analytics.  A functional comparison to approaches based on MapReduce (Section \ref{sec:mr}) has shown that DIMSpan is superior in terms of network traffic, disk access and the number of isomorphism resolutions. Our experimental evaluation showed the high scalability of DIMSpan for very large datasets and low support thresholds, not only on molecular data but also on directed multigraphs. 

We found that its runtime benefits most from optimized data structures as well as cheap and effective compression based thereon. In future work we will further optimize data representation and compression to count and exchange only as few bits as possible. Further on, we will investigate adaptive partitioning strategies to optimize load balancing among threads. 


\section{Acknowledgments}

This work is partially funded by the German Federal Ministry of Education and Research under project ScaDS Dresden/Leipzig (BMBF 01IS14014B).

\begin{small}
  \bibliographystyle{abbrv}
\bibliography{04_main} 
\end{small}


\end{document}